\RequirePackage{fix-cm}
\documentclass[smallextended]{svjour3New}       
\smartqed  
\usepackage{graphicx}
\usepackage{amsmath}
\usepackage{nicefrac}
\usepackage{slashed}
\usepackage{upgreek}
\usepackage{xcolor}
\usepackage{cite}
\usepackage{multirow}

\def\beq{\begin{equation}}
\def\eeq{\end{equation}}
\def\bea{\begin{eqnarray}}
\def\eea{\end{eqnarray}} 

\def\eqlab#1{\label{eq:#1}}
\def\figlab#1{\label{fig:#1}}

\def\seclab#1{\label{sec:#1}}
\def\eref#1{(\ref{eq:#1})}
\def\eqref#1{eq.~(\ref{eq:#1})}
\def\Eqref#1{Eq.~(\ref{eq:#1})}

\def\Figref#1{Fig.~\ref{fig:#1}}

\def\nn{\nonumber}
\def\dd{\mathrm{d}}

\DeclareMathOperator\im{Im}

\def\bq{\boldsymbol{q}}

\def\al{\alpha}
\def\be{\beta}
\def\ga{\gamma} \def\Ga{{\it\Gamma}}

\def\dd{{\rm d}}

\begin{document}

\title{$\Delta(1232)$-Resonance 
in the Hydrogen Spectrum
}


\author{Franziska Hagelstein
}


\institute{Franziska Hagelstein \at
Albert Einstein Center for Fundamental Physics\\
 Institute for Theoretical Physics -- University of Bern\\
 Sidlerstrasse 5, CH-3012 Bern, Switzerland\\
  \email{hagelstein@itp.unibe.ch} 
}

\date{\today}

\maketitle

\begin{abstract}
The electromagnetic excitation of the $\Delta(1232)$-resonance plays an appreciable role in the 
Lamb shift and hyperfine structure of muonic and electronic hydrogen.
Its effect appears at the subleading order $\mathcal{O}(\alpha^5)$, together with other proton-polarizability contributions from forward two-photon exchange.
We use the large-$N_c$ relations for the nucleon-to-delta transition form factors to compute
the effect of the $\Delta(1232)$ in the hydrogen spectrum. We pay particular attention to a subtile difference between predictions based on a direct calculation of the two-photon exchange (or Compton scattering amplitudes) \cite{Faustov:1998mc} and predictions based on the $\Delta(1232)$-production photoabsorption cross sections \cite{Buchmann:2009uqa}. The mismatch is explained by studying the dispersion relations for tree-level Compton scattering off the proton in more details.
\end{abstract}

\section{Introduction}
\label{intro}
\begin{figure}[tbh]
    \centering
    \begin{minipage}{.475\columnwidth}
        \centering
 \includegraphics[width=.8\columnwidth]{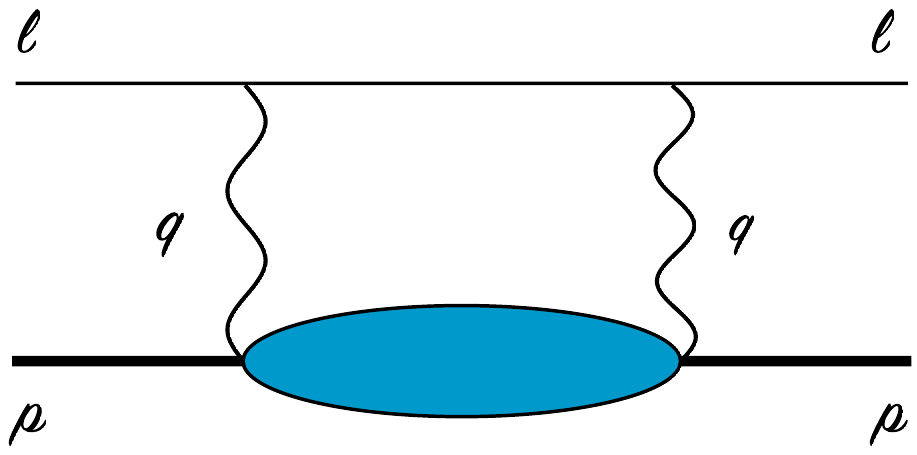}
\caption{Two-photon-exchange diagram in forward kinematics: the horizontal lines correspond to the lepton and the proton (bold). The ``blob'' represents all possible excitations in the non-Born diagrams.\label{fig:TPE}}
    \end{minipage}\hfill
    \begin{minipage}{.475\columnwidth}
        \centering\vspace{-0.25cm}
      \includegraphics[width=0.6\columnwidth]{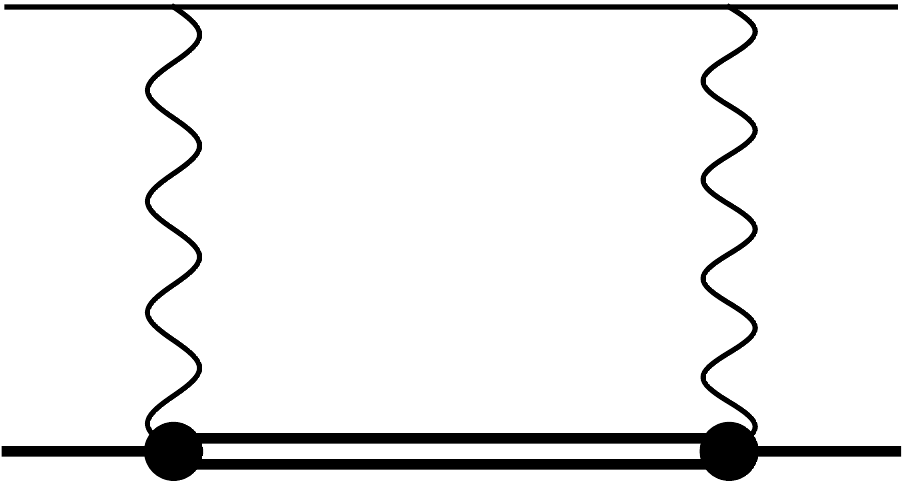}
       \caption{Two-photon-exchange diagram with intermediate $\Delta(1232)$-excitation. The crossed diagram is not drawn.}
              \label{fig:TPEDelta}
    \end{minipage}
\end{figure}

Spectroscopy of muonic hydrogen ($\mu$H) has great potential for precise extractions of proton structure informations, such as the proton charge radius. The $\mu$H Lamb shift experiment, performed by the CREMA collaboration \cite{Pohl:2010zza,Antognini:1900ns}, provided the currently most precise determination of the proton charge radius. Their value is about $10$ times more accurate than the CODATA average of experiments with electronic probes \cite{Mohr:2015ccw}, but $5.6\,\sigma$ smaller --- hence, posing the {\it proton radius puzzle}. Evidently, the extraction of the charge radius from the experimental Lamb shift or the Zemach radius from the measured hyperfine splitting (HFS), strongly depends on the quality of the theoretical input (summarized f.i.\ in Ref.~\cite{Antognini:2013rsa}). 
The biggest theoretical uncertainty comes from the forward two-photon exchange (TPE) between muon and proton, or rather, the proton-polarizability effect given by the non-Born contributions to the TPE, see \Figref{TPE}.
These effects are of the order $\mathcal{O}(\alpha^5)$, and therefore subleading with respect to the proton charge radius contribution which is of order $\mathcal{O}(\alpha^4)$.\footnote{See Ref.~\cite{Hagelstein2015} for a recent review on polarizabilities in Compton scattering and hydrogen.}

At present, the experimental information on the HFS in $\mu$H, used to extract the Zemach radius of the proton \cite{Antognini:1900ns,Antognini:2013rsa}, only comes from the $2S$ level. 
In the future, the planned measurements of the ground-state $1S$ HFS in $\mu$H (CREMA \cite{Pohl:2016xsr}, FAMU \cite{Adamczak:2016pdb} and J-PARC / Riken-RAL \cite{Sato:2014uza}) will improve the experimental HFS accuracy considerably, and thereby call for at least a factor of $10$ improvement in precision of the theory predictions of proton-polarizability effects \cite{Pohl:2016xsr}.

In this conference proceedings, we discuss the polarizability effect on the hydrogen spectrum generated by the $\Delta(1232)$-resonance through the diagram in \Figref{TPEDelta} (Lamb shift in Section \ref{sec:LS}, HFS in Section \ref{sec:HFS}).\footnote{The results have been previously presented in Ref.~\cite{Hagelstein:2017cbl}.} 
Thereby, our main aim is rather pedagogical, 
as we want to remind the reader of an interesting issue appearing in the tree-level Compton scattering (CS)  --- namely, the mismatch of Compton scattering amplitudes and dispersion relations with input from single-particle-production photoabsorption cross sections.

The theoretical framework is briefly presented in Section \ref{framework}.
In Section \ref{sec:CS}, we give a clarifying presentation of the tree-level CS process, with the Born diagrams discussed in Section \ref{Born}, and CS with intermediate $\Delta(1232)$ exchange calculated in Sections \ref{dif} and analyzed in Section \ref{disp}. A detailed study of the leading order (LO) plus $\Delta$  prediction of the $\al^5$-proton-polarizability contribution to the HFS in electronic and muonic hydrogen (H and $\mu$H) from baryon chiral perturbation theory (B$\chi$PT), discussing also the LO pion-cloud contribution, is postponed to Ref.~\cite{Hagelstein:2018}. B$\chi$PT studies of the Lamb shift at LO and LO plus $\Delta$ can be found in Refs.~\cite{Alarcon:2013cba} and \cite{Lensky:2017bwi}, respectively.  Further model-independent studies of the Lamb shift and the hyperfine splitting in H and $\mu$H can be found in Refs.~\cite{Pineda:2002as,Peset:2014jxa,Peset:2014yha,Peset:2016wjq}, which use the frameworks of heavy-baryon chiral perturbation theory (HB$\chi$PT) and non-relativistic QED.

\section{Theoretical Framework} \label{framework}
The forward TPE, shown in \Figref{TPE}, can be split into a leptonic and a hadronic tensor. The leptonic side can be calculated from QED, while the hadronic side is given by the amplitudes of forward doubly-virtual Compton scattering (VVCS) off the proton. The VVCS tensor splits into symmetric and antisymmetric parts:
\beq
T^{\mu \nu} (q,p) = \left[T^{\mu \nu}_S+T^{\mu \nu}_A\right] (q,p) ,
\eeq
which read:\footnote{Here and in the following, we use: $\ga_{\mu\nu}=\mbox{$\nicefrac{1}{2}$}\left[\ga_\mu,\ga_\nu\right]$, $\ga_{\mu\nu\al}=\mbox{$\nicefrac{1}{2}$}(\ga_\mu\ga_\nu\ga_\al - \ga_\al\ga_\nu\ga_\mu)$ and $\ga_{\mu\nu\al\be} = 
\mbox{$\nicefrac{1}{2}$}\left[\ga_{\mu\nu\al},\ga_\be\right]$.}
\begin{subequations}
\bea
T^{\mu \nu}_S(q,p) & = & -g^{\mu\nu}\,
T_1(\nu, Q^2)  +\frac{p^{\mu} p^{\nu} }{M^2} \, T_2(\nu, Q^2), \eqlab{VVCS_TS}\\
T^{\mu \nu}_A (q,p) & = &-\frac{1}{M}\gamma^{\mu \nu \al} q_\al \,S_1(\nu, Q^2) +
\frac{Q^2}{M^2}  \gamma^{\mu\nu} S_2(\nu, Q^2).
\eea
\end{subequations}
The spin-independent VVCS amplitudes, $T_1$ and $T_2$, contribute to the classic ($2P-2S$) Lamb shift \cite{Carlson:2011zd}:
\bea
\Delta E^\mathrm{TPE}(nS)&=& 8\pi \al m \,\phi_n^2\,
\frac{1}{i}\int_{-\infty}^\infty \!\frac{\dd\nu}{2\pi} \int \!\!\frac{\dd \bq}{(2\pi)^3}   \times\eqlab{VVCS_LS}\\
&&\times\frac{\left(Q^2-2\nu^2\right)T_1(\nu,Q^2)-(Q^2+\nu^2)\,T_2(\nu,Q^2)}{Q^4(Q^4-4m^2\nu^2)},\nn\qquad
\eea
whereas the spin-dependent VVCS amplitudes, $S_1$ and $S_2$, contribute to the HFS \cite{Carlson:2008ke}:
\bea
\frac{E^\mathrm{TPE}_{\mathrm{HFS}}(nS)}{E_\mathrm{F}(nS)}&=&\frac{4m}{1+\kappa}\frac{1}{i}\int_{-\infty}^\infty \!\frac{\dd\nu}{2\pi} \int \!\!\frac{\dd \bq}{(2\pi)^3}\times\eqlab{VVCS_HFS} \\
&&\times \frac{1}{Q^4-4m^2\nu^2}\left\{\frac{\left(2Q^2-\nu^2\right)}{Q^2}S_1(\nu,Q^2)+\frac{3\nu}{M}S_2(\nu,Q^2)\right\}.\nn
\eea
Here, $m$ is the muon mass, $M$ is the proton mass, $\kappa$ is the anomalous magnetic moment of the proton, $\nu$ is the photon energy in the lab frame, $q^2=-Q^2$ is the virtuality of the photon, $\phi_n^2=m_r^3 \al^3/(\pi n^3)$ is the hydrogen wave function of the $n$-th $S$-level at the origin, and $m_r$ is the reduced mass of the muon-proton system.

B$\chi$PT studies of the nucleon VVCS process can be found in Refs.~\cite{Bernard:2002pw,Bernard:2007zu,Bernard:2012hb} and Ref.~\cite{Lensky:2014dda}. These papers use the $\epsilon$- \cite{Hemmert:1996xg} and $\delta$-expansion \cite{Pascalutsa:2002pi}  power-counting schemes, respectively.\footnote{Note that in the large-$N_c$ limit, applied in Section \ref{sec:Jones}, the excitation energy of the $\Delta$ is vanishing \cite{Cohen:2002sd}: $\boldsymbol{\varDelta}=M_\Delta-M_N=\mathcal{O}(N_c^{-1})$. Therefore, in that limit the $\epsilon$ counting, which treats nucleon- and $\Delta$-propagators in the same way, is more appropriate. However, this difference is not affecting our calculation of the $\Delta$-exchange contribution. I thank Dr.~H.~Krebs for this remark.} 

To calculate the VVCS amplitudes of interest to this paper, we need the $\gamma^*N \rightarrow\Delta$ transition vertex:\footnote{The chiral Lagrangian for the $\gamma^*N \rightarrow\Delta$ interaction reads \cite{Pascalutsa:2005ts}:
\bea
\eqlab{nmGammaNDeltaLag}
 \mathcal{L}^{(2) \,\mathrm{nm}}_{\gamma N \Delta}&=&\frac{3e}{2M_N(M_N+M_\Delta)}\Bigg[\bar{N} T_3\Big\{i g_M (\partial_\mu \Delta_\nu) \tilde{F}^{\mu \nu}-g_E  \gamma_5 (\partial_\mu \Delta_\nu) F^{\mu \nu}\eqlab{LagrangianGND}\\
 &&+i \frac{g_\mathrm{C}}{M_\Delta}\gamma_5 \gamma^\al (\partial_\al \Delta_\nu-\partial_\nu \Delta_\al)\partial_\mu F^{\mu \nu}\Big\}+\Big\{g_E (\partial_\mu \bar\Delta_\nu)\gamma_5 F^{\mu \nu}\nn\\
 &&-i g_M (\partial_\mu \bar\Delta_\nu) \tilde{F}^{\mu \nu}+i \frac{g_\mathrm{C}}{M_\Delta}(\partial_\al \bar\Delta_\nu-\partial_\nu \bar\Delta_\al)\gamma^\al\gamma_5  \partial_\mu F^{\mu \nu}\Big\}T_3^\dagger N\Bigg],\nn
 \eea
where $N(x)$ and $\Delta_\mu(x)$ are the nucleon and Delta fields, and $T_3$ is an isospin 1/2 to 3/2 transition matrix \cite{Pascalutsa:2005vq,Ledwig:2011cx}.}

\hspace{-0.25cm}\begin{minipage}{2cm}{\centering\includegraphics[width=2.25cm]{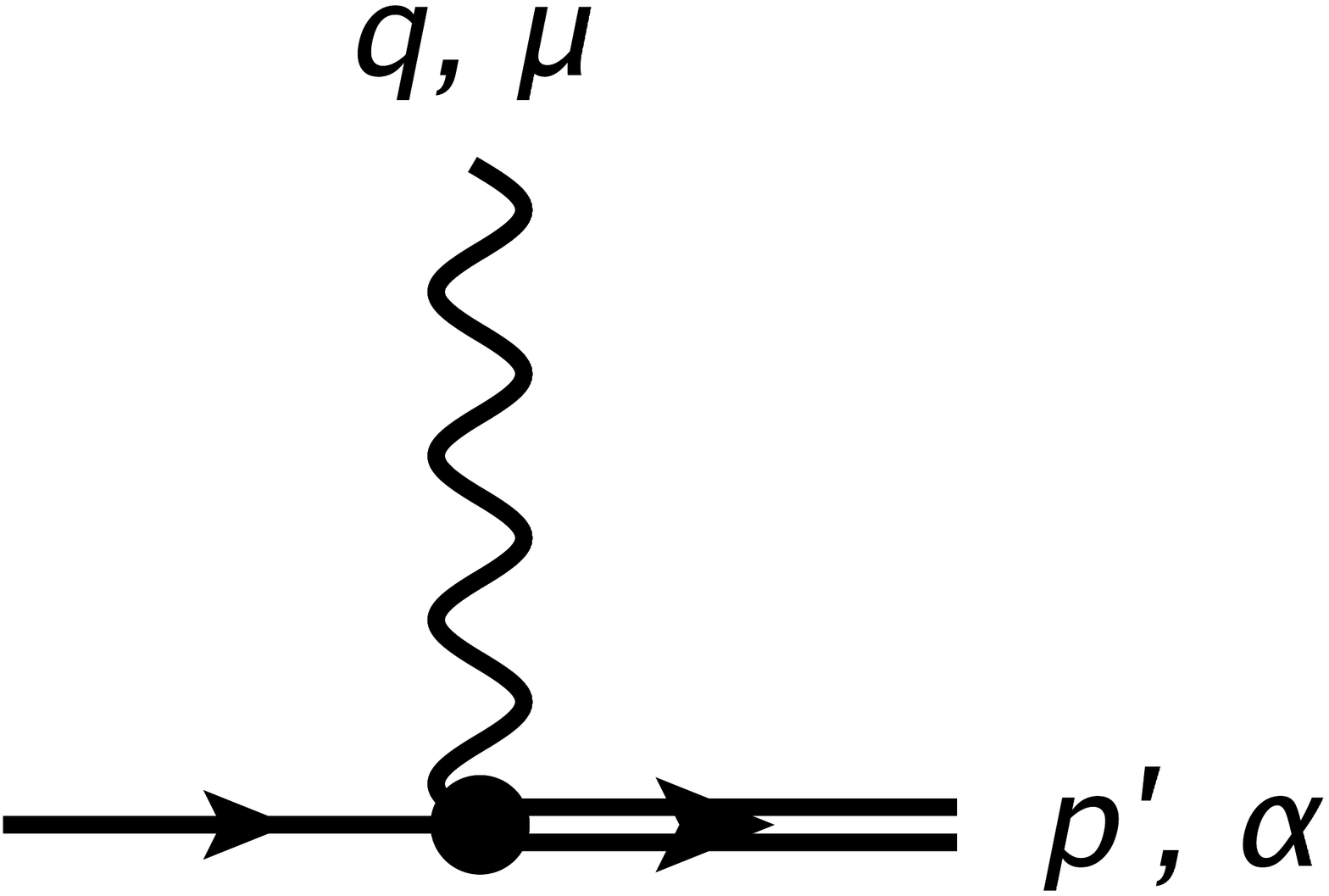}} 
\end{minipage}\hfill
\begin{minipage}{11cm}\begin{align}
\Ga_{\gamma N \rightarrow \Delta}^{\al \mu}(p',q)&=\sqrt{\frac{3}{2}}\frac{e}{M_N\left(M_N+M_\Delta\right)}\bigg\{g_M \gamma^{\al \mu \kappa \lambda}p'_\kappa q_\lambda\nn\\
&\quad+g_E (p'\cdot q\, g^{\al \mu}-q^\al p^{\prime\mu})+\frac{g_\mathrm{C}}{M_\Delta} \left(q^2 g^{ \al \mu}\slashed{p}'\right.\nn\\
&\quad\left.-q^2 p^{\prime\mu} \gamma^\al +p'\cdot q \,q^\mu \gamma^\al-q^\al q^\mu \slashed{p}'\right)\!\bigg\}\gamma_5,\qquad\nn
\end{align}\end{minipage}
and the spin-\mbox{$\frac{3}{2}$} propagator for the $\Delta(1232)$ \cite{Pascalutsa:2003aa}:

\hspace{-0.25cm}\begin{minipage}{2cm}{\includegraphics[width=2.25cm]{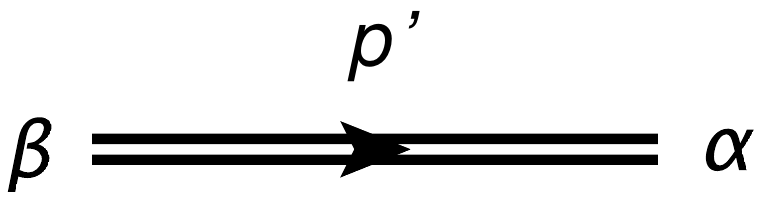}} 
\end{minipage}\hfill
\begin{minipage}{11cm}\begin{align}
S^{\al \be}_\Delta(p')&=\frac{\slashed{p}'+M_\Delta}{p^{\prime\,2}-M_\Delta^2+i 0^+}\left[-g^{\al \be}+\frac{1}{3} \gamma^\al \gamma^\be\right.\nn\\
&\left.\quad+\frac{1}{3M_\Delta}\left(\gamma^\al p^{\prime\,\beta}-\gamma^\beta p^{\prime\,\al}\right)+\frac{2}{3M_\Delta^2}p^{\prime\,\al}p^{\prime\,\beta}\right].\nn
\end{align}\end{minipage}

\vspace{\baselineskip}
\noindent Here, $M_N= 938.27$ MeV and $M_\Delta=1232$ MeV are the nucleon and Delta masses, and $g_M=2.97$, $g_E=-1.0$ and $g_\mathrm{C}=-2.6$ are the magnetic, electric and Coulomb couplings \cite{Pascalutsa:2005vq}.

\subsection{Jones--Scadron Form Factors and the Large-${N_c}$ Limit}\seclab{Jones}

As the VVCS amplitudes need to be integrated over the full range of $Q^2$ in Eqs.~\eref{VVCS_LS} and \eref{VVCS_HFS}, it is useful to relate our predictions derived from the above Feynman rules to empirical observables by means of large-$N_c$ relations \cite{tHooft:1973alw,Witten:1979kh}, thereby improving the convergence in $Q^2$. The magnetic ($g_M$), electric ($g_E$) and Coulomb ($g_\mathrm{C}$) couplings are per definition related to the magnetic $(G_M^*)$, electric $(G_E^*)$ and Coulomb $(G_\mathrm{C}^*)$ nucleon-to-delta transition form factors (FFs) of Jones and Scadron \cite{Jones:1972ky}:
\begin{subequations}
\eqlab{48}
\bea
g_M&=&G_M^*(Q^2)-G_E^*(Q^2),\\
g_E&=&-\frac{Q_+^2}{\omega_-^2+Q^2}\left[\frac{\omega_-}{M_\Delta}G_E^*(Q^2)+\frac{Q^2}{2M_\Delta^2}G_\mathrm{C}^*(Q^2)\right],\\
g_\mathrm{C}&=&\frac{Q_+^2}{\omega_-^2+Q^2}\left[G_E^*(Q^2)-\frac{\omega_-}{2M_\Delta}G_\mathrm{C}^*(Q^2)\right],
\eea
\end{subequations}
with 
\begin{subequations}
\eqlab{shorthands}
\bea
Q_+&=&\sqrt{(M_\Delta+ M)^2+Q^2}, \\
\omega_-&=&(M_\Delta^2-M^2- Q^2)/2M_\Delta.
\eea
\end{subequations}
These transition FFs are associated with the multipoles of pion electroproduction at the $\Delta(1232)$-resonance position, $M_{1+}^{(3/2)}$, $E_{1+}^{(3/2)}$ and $S_{1+}^{(3/2)}$, and enter the measured multipole ratios in the following way \cite{Pascalutsa:2007wz}:
\begin{subequations}
\eqlab{multipoles}
\bea
R_\mathrm{EM}(Q^2)&=&-\frac{G_E^*(Q^2)}{G_M^*(Q^2)},\eqlab{REM}\\
R_\mathrm{SM}(Q^2)&=&-\frac{Q_+Q_-}{4M_\Delta^2}\frac{G_\mathrm{C}^*(Q^2)}{G_M^*(Q^2)}.\eqlab{RSM}
\eea
\end{subequations}
They can be conveniently connected to the electromagnetic nucleon properties via large-$N_c$ relations:
\begin{subequations}
\bea
G_M^*(0)&=&\frac{\kappa_V}{\sqrt{2}} \quad \mbox{\cite{Jenkins:1994md}}, \\
G_E^*(0)&=&\frac{M^2-M_\Delta^2}{12\sqrt{2}}\left(\frac{M}{M_\Delta}\right)^{3/2}\langle r^2\rangle_{En} \quad\mbox{\cite{Buchmann:2002mm}},\eqlab{GE*0}\\
G_\mathrm{C}^*(0)&=&\frac{4M_\Delta^2}{M_\Delta^2-M^2}\,G_E^*(0)\quad \mbox{\cite{Pascalutsa:2007wz}},
\eea
\end{subequations}
where we introduced the isovector anomalous magnetic moment of the nucleon: $\kappa_V=\kappa_p-\kappa_n\simeq 3.7$ \cite{Olive:2016xmw}. An extension of these relations to finite momentum transfer is modeled and compared with available data for $R_\mathrm{EM}$ and $R_\mathrm{SM}$ in Ref.~\cite[Fig.~1]{Pascalutsa:2007wz}:
\begin{subequations}
\eqlab{PascalutsaLNC}
\bea
\eqlab{GM*old}
G_M^*(Q^2)&=&\frac{1}{\sqrt{2}}\left[F_{2p}(Q^2)-F_{2n}(Q^2)\right],\\
G_E^*(Q^2)&=&\left(\frac{M}{M_\Delta}\right)^{3/2}\frac{\varDelta M_+}{2\sqrt{2}\,Q^2}G_{En}(Q^2),\eqlab{GM*old2}\\
G_\mathrm{C}^*(Q^2)&=&\frac{4M_\Delta^2}{\varDelta M_+}G_E^*(Q^2).\eqlab{GM*old3}
\eea
\end{subequations}
Here, $F_{2p}$ and $F_{2n}$ are the Pauli FFs of the proton and neutron, respectively, while $G_{En}$ is the electric Sachs FF of the neutron. 

\begin{figure}[h!]
  \centering
 \includegraphics[scale=0.45]{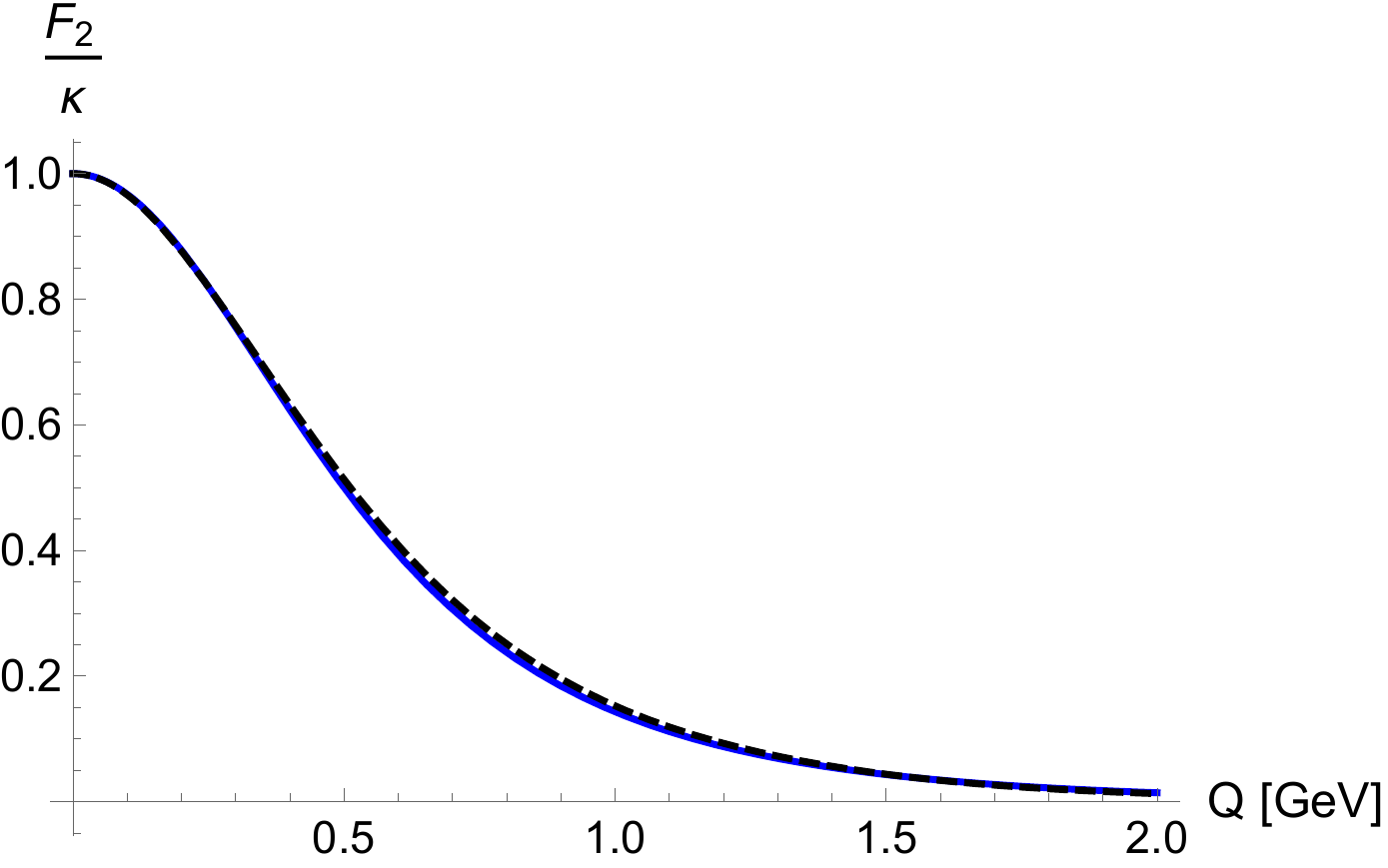}
  	\caption{Comparison of Pauli form factors normalized to the anomalous magnetic moment: blue solid line $F_{2p}(Q^2)/\kappa_p$, black dashed line $\left[F_{2p}(Q^2)-F_{2n}(Q^2)\right]/\kappa_V$.}  
	\figlab{Pauli}
\end{figure} 

Using the fact that $F_{2p}(Q^2)/\kappa_p\approx\left[F_{2p}(Q^2)-F_{2n}(Q^2)\right]/\kappa_V$, as illustrated in \Figref{Pauli}, one can further simplify \Eqref{GM*old}:
\beq
\eqlab{GM*}
G_M^*(Q^2)=\sqrt{2} \,C_M^* F_{2p}(Q^2),
\eeq
with $C_M^*=\frac{3.02}{\sqrt{2}\,\kappa_p}$ chosen such that the empirical value of $G_M^*(0)\simeq3.02$ \cite{Tiator:2003xr} is reproduced. In the following evaluation, we make the same choice as Ref.~\cite{Pascalutsa:2007wz} and apply the parametrizations of Ref.~\cite{Bradford:2006yz} for the electromagnetic nucleon FFs. 

\section{Compton Scattering off the Proton} \label{sec:CS}

In this section, we want to compare two different approaches to the tree-level CS. On one hand, we show a direct calculation of the VVCS amplitudes. On the other hand, we use CS sum rules with single-particle-production photoabsorption cross sections as input. This is a rather pedagogical discussion to remind the reader of a subtile difference between the two approaches.

\begin{figure}[tbh]
  \centering
  \includegraphics[scale=0.22]{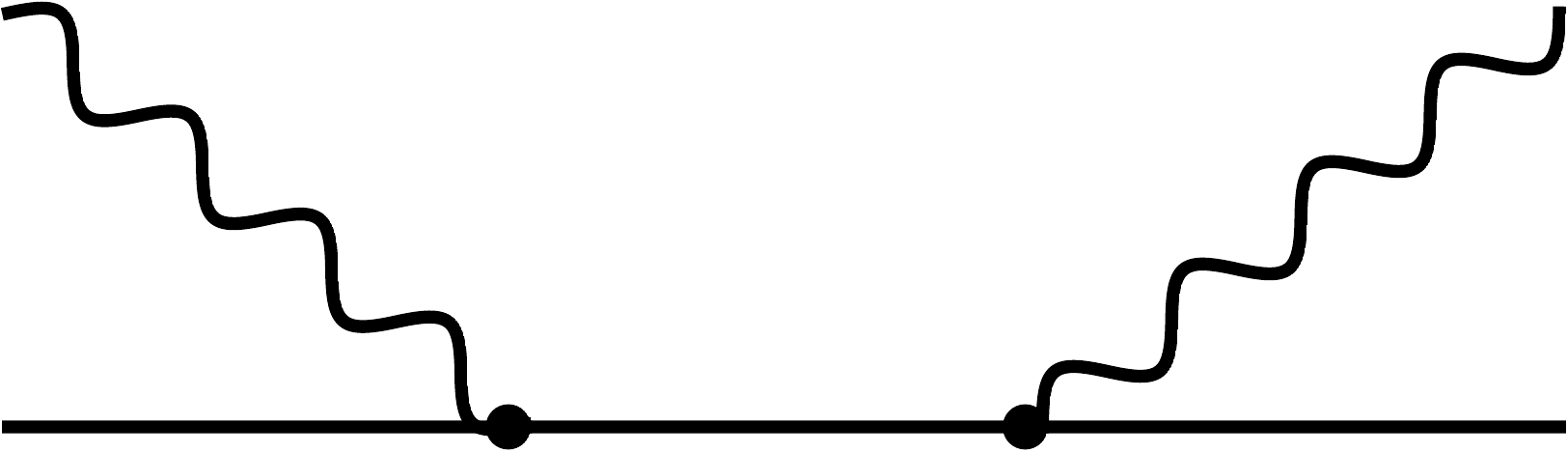}\hspace{0.75cm}
 \includegraphics[scale=0.22]{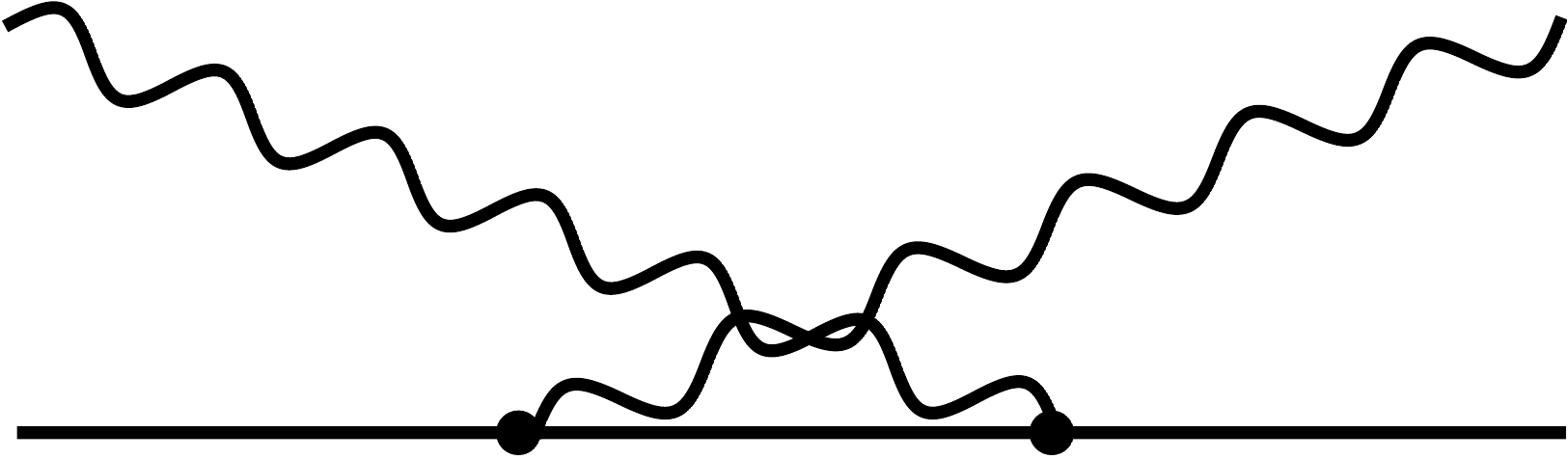}
  	\caption{Tree-level Compton scattering off the proton.}  
	\figlab{Born}
\end{figure} 

\subsection{Born Diagrams} \label{Born}

Let us first take a look at the leading diagrams in the process of CS off the proton, i.e., the tree-level Born diagrams shown in \Figref{Born}. The Born amplitudes are very well known \cite{Drechsel:2002ar}:\footnote{Since $S_2^\mathrm{Born}(\nu,Q^2)$ has a pole in the subsequent limits of $Q^2\rightarrow 0$ and $\nu\rightarrow 0$, we rather consider $\nu S_2^\mathrm{Born}(\nu,Q^2)$.}
\begin{subequations}
\bea
T_1^{\mathrm{Born}}(\nu, Q^2) &=&\frac{4\pi \alpha}{M}\left[-F_{1p}^2(Q^2)+\frac{\nu_{\mathrm{el}}^2\,  G_{Mp}^2(Q^2)}{\nu_{\mathrm{el}}^2-\nu^2
-i 0^+ }\right]\!, \\
T_2^{\mathrm{Born}}(\nu, Q^2)&=&\frac{8\pi \alpha\,  \nu_{\mathrm{el}} }{\nu_{\mathrm{el}}^2-\nu^2
-i 0^+ }\frac{G^{2}_{Ep}(Q^2) + \tau G^2_{Mp}(Q^2)}{1 + \tau},
\\
S_1^{\mathrm{Born}}(\nu, Q^2) &=& \frac{2\pi\al}{M}\left[-F_{2p}^2(Q^2)+\frac{2M \, \nu_{\mathrm{el}}}{\nu_{\mathrm{el}}^2-\nu^2
-i 0^+ } F_{1p}(Q^2)\, G_{Mp}(Q^2)\right]\!,\qquad\;\\
\nu S_2^{\mathrm{Born}}(\nu, Q^2)&=&2 \pi \alpha\, F_{2p}(Q^2)\, G_{Mp}(Q^2) \left[1+\frac{\nu_{\mathrm{el}}^2}{\nu_{\mathrm{el}}^2-\nu^2
-i 0^+ }\right]\! , \eqlab{S2subpole}
\eea
\end{subequations}
with the Dirac FF of the proton $F_{1p}$, and the electromagnetic Sachs FFs of the proton $G_{Ep}$ and $G_{Mp}$. As one can see, the terms containing $[\nu_{\mathrm{el}}^2-\nu^2
-i 0^+ ]^{-1}$ are complex valued ($0^+$ is an infinitesimally small positive number), they have a pole at the elastic threshold, $\nu_\mathrm{el}=\nicefrac{Q^2}{2M}$, and fulfil a dispersion relation by themselves. The elastic proton structure functions associated with these nucleon-pole terms ($T_i^{\mathrm{pole}}$ and $S_i^{\mathrm{pole}}$) read as:
\begin{subequations}
\bea
f_1^{\mathrm{el}}(x,Q^2) & =& \frac{1}{2}\,  G^2_M(Q^2)\,  \delta( 1 - x ), \\
f_2^{\mathrm{el}}(x,Q^2) & =& \frac{1}{1 + \tau} \, \big[G^2_E(Q^2) + \tau G^2_M(Q^2) \big]\, \delta( 1 - x), \\
g_1^{\mathrm{el}}(x,Q^2) & =&  \frac{1}{2}\,  F_1(Q^2) \,G_M(Q^2) \, \delta(1 - x), \\
g_2^{\mathrm{el}}(x,Q^2) & =& -  \frac{\tau}{2}\, F_2(Q^2)\, G_M(Q^2)\,  \delta(1 - x),
\eea
\end{subequations}
with $x=\nicefrac{Q^2}{2M\nu}$ the Bjorken variable and $x=1$ the elastic point.
What is interesting here, some Born amplitudes have additional contributions from the Dirac and Pauli FFs of the proton, which are not of pole type:\footnote{Note that the polarizabilities are defined through the non-Born amplitudes. Therefore, distinguishing between Born and pole pieces is crucial in the evaluation of the polarizability contribution to the hydrogen spectrum.}
\begin{subequations}
\bea
T_1^{\mathrm{Born}}(\nu, Q^2) &=&-\frac{4\pi \alpha}{M}F_{1p}^2(Q^2)+T_1^{\mathrm{pole}}(\nu, Q^2), \\
S_1^{\mathrm{Born}}(\nu, Q^2) &=& -\frac{2\pi\al}{M}F_{2p}^2(Q^2)+S_1^{\mathrm{pole}}(\nu, Q^2),\\
\nu S_2^{\mathrm{Born}}(\nu, Q^2) &=& 2\pi\al\,F_{2p}(Q^2)G_{Mp}(Q^2)+\left[\nu S_2\right]^{\mathrm{pole}}(\nu, Q^2).
\eea
\end{subequations}
If one would want to describe these additional pieces through (unsubtracted) dispersion relations, structure functions proportional to $\delta(x)$ would be needed.
In the following, we will see that this separation of the tree-level amplitudes into \textit{pole} and \textit{non-pole pieces} is not unique for the Born diagrams, but also appears for the $\Delta(1232)$ as one-particle intermediate state.

\subsection{Tree-Level Compton Scattering with $\Delta(1232)$ Exchange} \label{dif}

\begin{figure}[tbh]
\centering
       \includegraphics[scale=0.35]{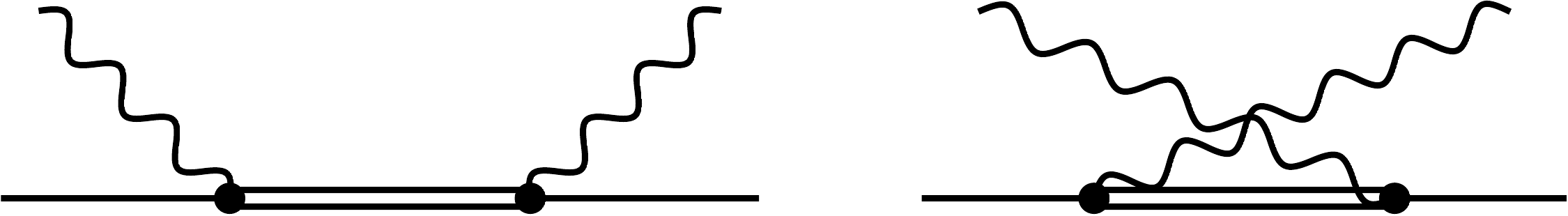}
\caption{$\Delta(1232)$-exchange contribution to tree-level Compton scattering off the proton.
\label{fig:DeltaExchange}}
\end{figure}

\begin{figure}[tb]
    \centering 
  \includegraphics[width=2.25cm]{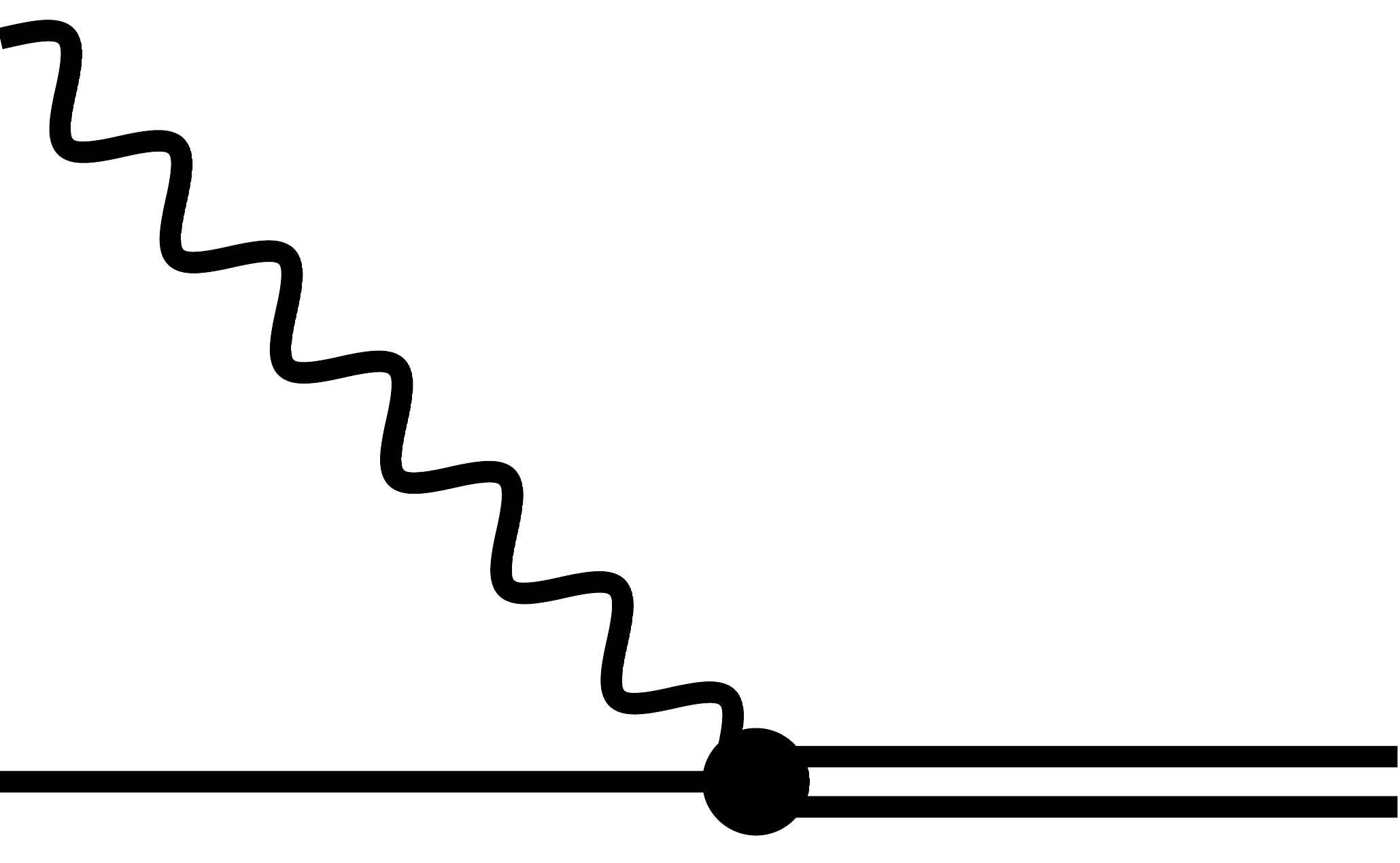}
\caption{Tree-level $\Delta(1232)$-production photoabsorption diagram.\label{fig:CrossSection}}
\end{figure}

The $\Delta(1232)$-exchange diagrams, shown in \Figref{DeltaExchange}, contribute to the nucleon polarizabilities at NLO in B$\chi$PT. Here, we present our results for the VVCS amplitudes and the $\Delta(1232)$-production photoabsorption cross sections, shown in \Figref{CrossSection}, in terms of the couplings $g_M$, $g_E$ and $g_C$. For brevity, we won't show analytic results after the substitution of the Jones--Scadron transition FFs, described in Section \ref{sec:Jones}.

The amplitudes for tree-level VVCS with $\Delta(1232)$ exchange can be decomposed in the following way:
\begin{subequations}
\eqlab{DeltaAmps}
\bea
T_1(\nu,Q^2)&=&T_1(0,Q^2)+T_1^{\Delta\mathrm{-pole}}(\nu,Q^2)+\widetilde T_1(\nu,Q^2)+\frac{i\,4\pi^2 \al}{M}\, f_1(\nu,Q^2),\qquad\;\\
T_2(\nu,Q^2)&=&T_2^{\Delta\mathrm{-pole}}(\nu,Q^2)+\widetilde T_2(\nu,Q^2)+\frac{i\,4\pi^2 \al}{\nu} \,f_2(\nu,Q^2),\\
S_1(\nu,Q^2)&=&S_1^{\Delta\mathrm{-pole}}(\nu,Q^2)+\widetilde S_1(\nu,Q^2)+\frac{i\,4\pi^2 \al}{\nu} \,g_1(\nu,Q^2),\\
\nu S_2(\nu,Q^2)&=&\nu S_2^{\Delta\mathrm{-pole}}(\nu,Q^2)+\widetilde{\nu S_2}(\nu,Q^2)+ \frac{i\,4\pi^2 \al M}{\nu} \,g_2(\nu,Q^2),
\eea
\end{subequations}
with the individual terms explained in what follows. The imaginary parts of the amplitudes are related to the unpolarized structure functions ($f_1$ and $f_2$) and the spin structure functions ($g_1$ and $g_2$) by the optical theorem:
\begin{subequations}
\eqlab{VVCSunitarity}
\bea
\im T_1(\nu,Q^2)&=&\frac{4\pi^2 \al }{M}f_1(x,Q^2)=K\,\sigma_T(\nu,Q^2), \eqlab{ImT1} \\
\im T_2(\nu,Q^2)&=&\frac{4\pi^2 \al}{\nu}f_2(x,Q^2)=\frac{Q^2 K}{\nu^2+Q^2}\left[\sigma_T+\sigma_L\right](\nu,Q^2), \eqlab{ImT2}\\
\im S_1(\nu,Q^2) &=& \frac{4\pi^2 \alpha}{\nu} \, g_1(x,Q^2) = 
\frac{M  K \nu}{\nu^2+Q^2}\left[\frac{Q}{\nu}\sigma_{LT}  + \sigma_{TT}\right](\nu,Q^2), \eqlab{ImS1}\\
\im S_2(\nu,Q^2) & =&  \frac{4\pi^2 \alpha M}{\nu^2} \, g_2(x, Q^2)  
= \frac{M^2 K}{\nu^2+Q^2}\left[\frac{\nu}{Q}\sigma_{LT}  - \sigma_{TT}\right](\nu,Q^2), \qquad \quad\eqlab{ImS2}
\eea
\end{subequations}
with the photon flux factor $K$. The analytic expressions for the $\Delta(1232)$-production structure functions are given in Appendix \ref{PhotoCS}.
The right-hand side of \Eqref{VVCSunitarity} shows how the proton structure functions are in turn related to the $\Delta(1232)$-production photoabsorption cross sections.\footnote{The cross sections are the usual combinations of helicity cross sections: $\sigma_T=\nicefrac12\, (\sigma_{1/2}+\sigma_{3/2})$ and $\sigma_{TT}=\nicefrac12\, (\sigma_{1/2}-\sigma_{3/2})$ for transversely polarized photons, and $\sigma_L=\nicefrac12\, (\sigma_{1/2}+\sigma_{-1/2})$ for longitudinal photons. The cross section $\sigma_{LT}$ 
describes a simultaneous helicity
change of the photon (from longitudinal to transverse) and the nucleon (spin-flip) such that the
total helicity is conserved. } The threshold for production of the $\Delta(1232)$-resonance is at lab-frame photon energies of:
\beq
\nu_\Delta=\frac{M_\Delta^2-M^2+Q^2}{2M}.
\eeq
Hence, the response functions are expected to be proportional to $\delta\!\left(\nu-\nu_\Delta\right)$.\footnote{Equivalently, we can write the $\delta$-function as:
\beq
\delta\! \left(\nu-\nu_\Delta\right)=\frac{Q^2}{2M\nu_\Delta^2}\,\delta\!\left(x-\nicefrac{Q^2}{2M\nu_\Delta}\right). \eqlab{dfunc}
\eeq} 

For the real parts of the amplitudes, given in Appendix \ref{amps}, we distinguish two kinds of structures.
Terms which are proportional to:
\beq
\frac{1}{[s-M_\Delta^2][u-M_\Delta^2]}=\frac{1}{4M^2}\frac{1}{\nu_\Delta^2-\nu^2}, \eqlab{poleStruc}
\eeq
are denoted as \textit{$\Delta$-pole terms}: $T_i^{\Delta\mathrm{-pole}}$ and $S_i^{\Delta\mathrm{-pole}}$, see \Eqref{Dpole}, where $s$ and $u$ are the usual Mandelstam variables. They have a pole at the $\Delta(1232)$-production threshold, i.e.\ at $\nu=\nu_\Delta$. In addition, we find the \textit{($\Delta$-)non-pole terms}, $\widetilde T_i$ and $\widetilde S_i$, which are free of poles in $\nu$, see \Eqref{noDpole}. They emerge as:
\beq
\frac{\nu^{n+2}}{[s-M_\Delta^2][u-M_\Delta^2]}=\frac{1}{4M^2} \left(\frac{\nu_\Delta^2 \nu^n}{\nu_\Delta^2-\nu^2}-\nu^n\right)\!,
\eeq
where in the second term the $\Delta$-pole has canceled out. 

As we can see, the pole terms, when derived from a local Lagrangian, can be accompanied by non-pole terms. 
This means that the pole terms by themselves do not necessarily satisfy all the general constraints, as we see happening, f.i., 
for the $\Delta$-pole term which by itself violates the BC sum rule, cf.\ discussion before \Eqref{BCSR}.

 \subsection{Dispersive Description of $\Delta(1232)$-Pole and Non-Pole Contributions}\label{disp}

It is important to emphasize that the response functions in \Eqref{structurefunc}, which are proportional to $\delta\!\left(\nu-\nu_\Delta\right)$, describe the production of a real $\Delta(1232)$ in the final state of the cross section. Therefore, as has been observed before, they can not reproduce the non-pole contributions to the VVCS amplitudes stated in \Eqref{noDpole}. However, the structure functions \eref{structurefunc} reproduce the $\Delta$-pole parts of the VVCS amplitudes, given in \Eqref{Dpole}, as we verified exploiting the dispersion relations (DRs): 
\begin{subequations}
\eqlab{DRVVCS}
\bea
\eqlab{T1DR}
T_1 ( \nu, Q^2) &=& T_1(0,Q^2) +\frac{32\pi\al M\nu^2}{Q^4}\int_{0}^1 
\,\dd x \, 
\frac{x f_1 (x, Q^2)}{1 - x^2 (\nu/\nu_{\mathrm{el}})^2 - i 0^+}\eqlab{T1Subtr},\quad\\
\eqlab{T2DR}
T_2 ( \nu, Q^2) &=& \frac{16\pi \al M}{Q^2} \int_{0}^1 
\!\dd x\, 
\frac{f_2 (x, Q^2)}{1 - x^2 (\nu/\nu_{\mathrm{el}})^2  - i 0^+},\\
\eqlab{S1DR}
S_1 ( \nu, Q^2) &=& \frac{16\pi \al M}{Q^2} \int_{0}^1 
\!\dd x\, 
\frac{g_1 (x, Q^2)}{1 - x^2 (\nu/\nu_{\mathrm{el}})^2  - i 0^+},\\
\eqlab{S2DR}
\nu S_2 ( \nu, Q^2) &=&  \frac{16\pi \al M^2}{Q^2} \int_{0}^1 
\!\dd x\, 
\frac{g_2 (x, Q^2)}{1 - x^2 (\nu/\nu_{\mathrm{el}})^2  - i 0^+}.
\eea
\end{subequations}
 Note that the high-energy asymptotics of $f_1$ require a once-subtracted dispersion relation for the $T_1$ amplitude, with the subtraction function $T_1(0,Q^2)$ written in \Eqref{T1su}.
 
 To describe the non-pole contributions in a dispersive framework, we introduce the following structure functions:
\begin{subequations}
\eqlab{npstrucfunc}
\bea
\widetilde f_1(x,Q^2)&=&\frac{M x}{8 \pi \al} \,\widetilde T_1(x,Q^2)\, \delta (x),\\
\widetilde f_2(x,Q^2)&=&\frac{Q^2}{16 \pi \al M} \,\widetilde T_2(x,Q^2)\, \delta (x),\\
\widetilde g_1(x,Q^2)&=&\frac{Q^2}{16 \pi \al M} \,\widetilde S_1(x,Q^2)\, \delta (x),
\eea
\end{subequations}
which reproduce Eqs.~\eref{T1nonpole}-\eref{S1nonpole} as plugged into the DRs in Eqs.~\eref{T1Subtr}-\eref{S1DR}, respectively. The non-pole part of $\nu S_2$ is a bit more complicated to reconstruct, because it has terms constant in $\nu$ and terms proportional to $\nu^2$. The $\nu$-independent part of \Eqref{NUnonpoleS2} can be described by: 
\beq
\widetilde g_{2,a}(x,Q^2)=\frac{Q^2 }{16 \pi  \al M^2}\Bigg[ \widetilde{\nu S_2}\Big\vert_{\nu\rightarrow0}\Bigg] \delta (x), \eqlab{S2npg}
\eeq
as plugged into \Eqref{S2DR}. The part of \Eqref{NUnonpoleS2} proportional to $\nu^2$ can be described based on:
\beq
\widetilde g_{2,b}(x,Q^2)=\frac{Q^6 }{64 \pi \al M^4} \frac{1}{x^2}\left[\frac{\widetilde S_2(Q^2)}{\nu}\right] \delta (x),\eqlab{S2npgp2}
\eeq
with the once-subtracted dispersion relation:
\beq
\widetilde{\nu S_2}(\nu,Q^2)-\widetilde{\nu S_2}(0,Q^2)=\frac{64 \pi \al M^4\nu^2}{Q^6}  \int_{0}^{x_0} 
\!\dd x\,
\frac{x^2 \,\tilde g_{2,b} (x, Q^2)}{1 - x^2 (\nu/\nu_{\mathrm{el}})^2}.
\eeq

In this section, we have shown that the amplitudes for tree-level CS with intermediate $\Delta(1232)$ can be split into a {\it $\Delta$-pole} and a {\it ($\Delta$-)non-pole} part. While DRs with $\Delta(1232)$-production cross sections are able to reproduce the $\Delta$-poles terms. They are obviously unable to give us the non-pole terms. To reconstruct these from DRs, we had to define some auxiliary response functions, cf.\ Eqs.~\eref{npstrucfunc}-\eref{S2npgp2}. It is very crucial to keep this observation in mind, as one should not omit the non-pole terms. We will show this explicitly when studying the TPE in \Figref{TPEDelta}.
Therefore, it is favourable to calculate the effect of the $\Delta(1232)$-resonance in the hydrogen spectrum based on the VVCS amplitudes, and not based on the $\Delta(1232)$-production cross sections. We will do so in the following Sections \ref{sec:LS} and \ref{sec:HFS} for the Lamb shift and the HFS.

\section{$\Delta(1232)$ in the Lamb Shift} \label{sec:LS}

The magnetic dipole polarizability is suppressed in the Lamb shift, as one can see from the low-energy expansion of the spin-independent VVCS amplitudes entering \Eqref{VVCS_LS}, see discussion in Ref.~\cite[Eq.~(12)]{Alarcon:2013cba}. Since the nucleon-to-delta transition is dominantly of magnetic dipole type, we expect the $\Delta(1232)$-resonance to have a numerically small influence on the Lamb shift.

As outlined in Section \ref{framework}, we calculate the TPE in \Figref{TPEDelta} based on the large-$N_c$ limit of the Jones-Scadron FFs extended to finite momentum transfers, i.e., Eqs.~\eref{GM*old2}, \eref{GM*old3} and \eref{GM*} with $C_M^*=\frac{3.02}{\sqrt{2}\,\kappa_p}$. For the empirical input, we rely on parametrizations of the elastic nucleon FFs \cite{Bradford:2006yz}. 

Our results are summarized in Table \ref{nonpoleLS}, where we present the contributions to the $2S$-level shift in $\mu$H from the subtraction function $T_1(0,Q^2)$, the $\Delta$-pole amplitudes $T_i^{\Delta\mathrm{-pole}}$ and the non-pole amplitudes $\widetilde T_i$ separately. The size of these individual contributions is comparable to the leading effect of  chiral dynamics~\cite{Alarcon:2013cba}:
\begin{equation}
\eqlab{LOLS}
E_\text{LS}^{\langle\text{LO}\rangle\,\mathrm{pol.}}(\mu \text{H})=8^{+3}_{-1}\, \upmu\mathrm{eV}.
\end{equation}
If we, however, combine all contributions, the $\Delta$-pole parts of the VVCS amplitudes largely cancel the subtraction function and the non-pole parts, cf.\ last column in Table \ref{nonpoleLS}. The resulting total effect of the $\Delta(1232)$ on the $2P_{1/2}-2S_{1/2}$ Lamb shift in $\mu$H then amounts to:
\beq
E_\text{LS}^{\langle\Delta\rangle\,\mathrm{pol.}}(\mu \text{H})=-0.95\pm0.95\, \upmu\mathrm{eV}. \eqlab{DeExLS}
\eeq
As was expected from the suppression of the magnetic dipole polarizability in the Lamb shift, the NLO $\al^5$-proton-polarizability contribution of the $\Delta$, \Eqref{DeExLS}, is substantially smaller than the leading $\al^5$-proton-polarizability effect, \Eqref{LOLS}. This result agrees with the model-independent calculation of Ref.~\cite[Eq.~(4.23)]{Peset:2014jxa} within errors.

To estimate the quality of our prediction, we confirmed that the contribution from large momentum transfers ($Q>m_\rho$) is less than $1\,\%$.
Furthermore, we checked that using a dipole FF for $G_{Ep}$ and $G_{Mp}$, as well as the Galster parametrization for $G_{En}$ \cite{Galster:1971kv}, or the Ramalho \cite{Ramalho:2017xkr} parametrization, which includes $\gamma^*N \rightarrow \Delta(1232)$ quadrupole FF data, for $G_{En}$ and the Bradford parametrization for $F_{2p}$ changes \Eqref{DeExLS} by less than $6\,\%$. Nevertheless, due to the discussed strong cancelations in the final result, shrinking it by one order of magnitude compared to f.i.\ the pure $\Delta$-pole effect, we assigned a conservative error of $100\,\%$ on our prediction in \Eqref{DeExLS}.

Let us now take a closer look at the effect of the subtraction function $T_1(0,Q^2)$. It is of special interest, since it is not known from experiment, and thus, has to be modeled in any ``data-driven'' dispersive approach to the TPE, cf.\ Refs.~\cite{Pachucki:1999zza,Martynenko:2005rc,Carlson:2011zd,Birse:2012eb,Gorchtein:2013yga}.
Surprisingly, the $\Delta$-exchange contribution to the subtraction term:
\beq
E_\text{LS}^{\langle\Delta\rangle\,\mathrm{subtr.}}(\mu \text{H})=-7.58\pm 2.27\, \upmu\mathrm{eV},\eqlab{deexsub}
\eeq
is much larger than the LO B$\chi$PT contribution from the $\pi N$ loops \cite{Alarcon:2013cba}:
\beq
E_\text{LS}^{\langle \text{LO}\rangle\,\mathrm{subtr.}}(\mu \text{H})=3^{+0.9}_{-0.5}\, \upmu\mathrm{eV}.
\eeq
Note that for the $\Delta$-exchange contribution in \Eqref{deexsub}, we assigned a $30\,\%$ error due to higher orders in the chiral expansion.
Therefore, it has a substantial effect on our prediction for the subtraction term  \cite{Lensky:2017bwi}:
\begin{equation}
E_\text{LS}^{\langle\text{LO+$\Delta$}\rangle\,\text{subtr.}}(\mu \text{H})=-4.6^{+2.4}_{-2.3}\, \upmu\mathrm{eV},
\end{equation}
which is in good agreement with the dispersive predictions from Refs.~\cite{Carlson:2011zd,Birse:2012eb}. A HB$\chi$PT prediction of the subtraction term, including leading and subleading $\pi$- and  $\pi\Delta$-loops, respectively,  can be found in Ref.~\cite{Peset:2014jxa}. However, the LO B$\chi$PT and HB$\chi$PT predictions of the Lamb shift and the subtraction term deviate, as is discussed in Ref.~\cite{Alarcon:2013cba}.

 \renewcommand{\arraystretch}{1.5}
\begin{table} [t]
\caption{$\Delta(1232)$-exchange contribution to the $2S$-level shift in muonic hydrogen. All values in $\upmu\mathrm{eV}$. \label{nonpoleLS}}
\centering
\begin{small}
\begin{tabular}{|c|c|c|c|}
\hline
{\bf Input}&$\boldsymbol{\Delta E(T_1)}$&$\boldsymbol{\Delta E(T_2)}$&$\boldsymbol{\Delta E}$\\
\hline
$T_1(0,Q^2)$ \eref{T1su}&$7.58$&/&$7.58$\\
 $T_i^{\Delta\mathrm{-pole}}$ \eref{Dpole}&$-2.22$&$-6.01$&$-8.23$\\
 $\widetilde T_i$ \eref{noDpole}&$0.40$&$1.19$&$1.59$\\
\hline
 $T_i$ \eref{DeltaAmps}&$5.76$&$-4.82$&$0.95$\\
\hline
\end{tabular}
\end{small}
\end{table}
 \renewcommand{\arraystretch}{1.3}

\section{$\Delta(1232)$ in the Hyperfine Splitting} \label{sec:HFS}

In the following, we study the effect of the $\Delta(1232)$-resonance on the HFS. The calculation proceeds analogously to the Lamb shift case, and as we will see in the subsequent discussion, a similar calculations can be found in the literature \cite{Buchmann:2009uqa}.\footnote{A first paper on the leading chiral logarithms in the HFS of H and $\mu$H, studying also the large-$N_c$ limit of the polarizability contribution, can be found in Ref.~\cite{Pineda:2002as}.}

Our results are summarized in Table \ref{nonpoleHFS}, where we show contributions to the $2S$ HFS in $\mu$H from the $\Delta$-pole amplitudes $S_i^{\Delta-\text{pole}}$, the non-pole amplitudes $\widetilde S_i$, and their combination. Just as in the Lamb shift calculation, the $\Delta$-pole and non-pole contributions largely cancel each other, hence, we assign a conservative $100\,\%$ error on their sum.
However, contrary to the Lamb shift situation, we find that the effect of the $\Delta(1232)$ on the $\mu$H HFS: 
\beq
E_{\text{HFS}}^{{\langle\Delta\rangle}\,\mathrm{pol.}}(2S, \mu\text{H})=-1.15\pm 1.15\,\upmu\mathrm{eV},\eqlab{HFSmuH}
\eeq
is certainly relevant in comparison to the leading chiral loops:
\beq
E_\mathrm{HFS}^{{\langle\text{LO}\rangle}\,\mathrm{pol.}}(2S, \mu\mathrm{H})=0.85^{\,+0.85}_{\,-1.08}\,\upmu\mathrm{eV}.\eqlab{pionresultHFS2SmuH}
\eeq

Again, we verified that the contribution from $Q<m_\rho$ is small ($5\,\%$) and that the result is not sensitive to the choice of a nucleon FF parametrization, i.e, using different parametrizations for the elastic nucleon FFs (dipole and Galster FFs \cite{Galster:1971kv}, or the Ramalho and Bradford parametrizations \cite{Bradford:2006yz,Ramalho:2017xkr}) leads to a change of less than $7\,\%$ in \Eqref{HFSmuH}.

Apart from the obvious ``quantitative'' argument that the non-pole terms should not be neglected, because they have a sizeable numerical effect on the HFS, we can give another, more ``qualitative'' explanation. Even though, $[\nu S_2]^{\Delta-\text{pole}}\vert_{\nu=0}$ and $\widetilde{\nu S_2}\vert_{\nu=0}$ both give contributions to the HFS, they cancel each other exactly. This is an important observation, because it means that only the combination of $\Delta$-pole and ($\Delta$-)non-pole pieces fulfils the Burkhardt-Cottingham (BC) sum rule \cite{Burkhardt:1970ti}:
\beq
\eqlab{BCSR}
\lim_{\nu \rightarrow 0}\;\frac{\nu S_2(\nu,Q^2)}{8\pi \al}=\frac{2M^2}{Q^2}\int_0^1 \dd x\, g_2(x,Q^2)=0.
\eeq
In other words, neglecting the non-pole terms would violate the BC sum rule and is therefore not a good approximation. I should note in passing that in Table \ref{nonpoleHFS} we used \Eqref{VVCS_HFS} with the BC sum rule removed. Including the BC sum rule [i.e.,   the $\nu S_2\vert_{\nu=0}$ amplitudes, or equivalently, the $0$\textsuperscript{th} moments of the structure functions in Eqs.~\eref{S2npg} and \eref{g2Delta}], the $S_2^{\Delta\mathrm{-pole}}$ ($\widetilde S_2$) contribution to the $2S$ HFS in $\mu$H increases (decreases) by $52.75\,\upmu$eV.

\renewcommand{\arraystretch}{1.5}
\begin{table} [t]
\centering
\begin{small}
\caption{$\Delta(1232)$-exchange contribution to the $2S$ hyperfine splitting in muonic hydrogen. All values in $\upmu\mathrm{eV}$. \label{nonpoleHFS}}
\centering
\begin{tabular}{|c|c|c|c|}
\hline
{\bf Input}&$\boldsymbol{E_\mathrm{HFS}(S_1)}$&$\boldsymbol{E_\mathrm{HFS}(S_2)}$&$\boldsymbol{E_\mathrm{HFS}}$\\
\hline
$S_i^{\Delta\mathrm{-pole}}$\eref{Dpole}&$-38.27$&$-2.43$&$-40.69$\\
$\widetilde S_i$ \eref{noDpole}&$39.53$&$0.02$&$39.54$\\
\hline
$S_i$ \eref{DeltaAmps}&$1.26$&$-2.41$&$-1.15$\\
\hline
\end{tabular}
\end{small}
\end{table}
\renewcommand{\arraystretch}{1.3}

To compare with the literature, let us now switch from muonic to normal, electronic hydrogen (H), where we find:
\beq
E_{\mathrm{HFS}}^{{\langle\Delta\rangle}\,\mathrm{pol.}}(2S,\mathrm{H})=-0.212\pm 0.212\,\mathrm{peV}.
\eeq
We define auxiliary quantities ($\Delta_\mathrm{pol.}$, $\delta_1$ and $\delta_2$) reflecting the size of the hyperfine structure independent of the hydrogen level under consideration:
\beq
E_\mathrm{HFS}^\mathrm{pol}(nS)=E_F(nS) \,\Delta_\mathrm{pol},\qquad \text{with}\;\;\Delta_\mathrm{pol}=\delta_1+\delta_2,
\eeq
where $E_F$ is the Fermi energy and $\delta_i$ corresponds to the contribution of $S_i$ or $g_i$, respectively. 

The effect of the $\Delta(1232)$-resonance on the HFS in H is summarized in Table \ref{Comparison}. There, we distinguish not only $\Delta$-pole and non-pole contributions, but also contributions from the different Jones--Scadron transition FFs, or equivalently, the different multipoles of pion electroproduction at the resonance position. 
If we look at the individual contributions from $\Delta$-pole and non-pole amplitudes, the HFS is unsurprisingly dominated by the magnetic dipole transition. Due to their large cancelation into the total result, this dominance is however weakened, and $G_\mathrm{C}^*$ and $G_E^*$ gain more impact. Therefore, considering only the magnetic nucleon-to-delta transition, represented by $G_M^{*2}$, is an unsatisfactory approximation for the HFS.

As one can see from Table \ref{Comparison}, our results compare well with Ref.~\cite{Buchmann:2009uqa} and Ref.~\cite{Faustov:1998mc}. The approach of Ref.~\cite{Buchmann:2009uqa} is very similar to the work presented in here, since it also uses large-$N_c$ relations for the nucleon-to-delta transition FFs, cf.\ Section \ref{framework}. However, Ref.~\cite{Buchmann:2009uqa} makes use of DRs for the VVCS amplitudes with the theoretical $\Delta(1232)$-production cross sections as input. In this way, it matches the $\Delta$-pole contribution, but misses the non-pole contributions as explained in Section \ref{disp}. On the other hand, Ref.~\cite{Faustov:1998mc} calculates the TPE directly, with input from experimental data on nucleon-to-delta transition FFs. Hence, it in principle obtains the whole effect of the $\Delta(1232)$, however, with $G_E^*$ and $G_\mathrm{C}^*$ neglected.

\renewcommand{\arraystretch}{1.5}
\begin{table} [tb]
\centering
\begin{small}
\caption{Comparison of different predictions for the $\Delta(1232)$-exchange contribution to the $2S$ hyperfine splitting in electronic hydrogen.\label{Comparison}}
\centering
\begin{tabular}{|c|c|c|c|c|c|c|c|}
\hline
\multicolumn{2}{|c|}{\bf Type of contribution}&\multicolumn{2}{|c|}{\bf $\boldsymbol{\delta_1}$ [ppm]}&\multicolumn{2}{|c|}{\bf $\boldsymbol{\delta_2}$ [ppm]}&\multicolumn{2}{|c|}{\bf $\boldsymbol{\Delta_\mathrm{pol}}$ [ppm]}\\
\hline
\hline
\multirow{ 3}{*}{\rotatebox[origin=c]{90}{$S_i^{\Delta\mathrm{-pole}}$}}&all multipoles&\multicolumn{2}{|c|}{$-34.82$}&\multicolumn{2}{|c|}{$-0.71$}&\multicolumn{2}{|c|}{$-35.53$}\\
\cline{3-8}
&$G_M^{*2}$&\multicolumn{2}{|c|}{$-31.27$}&$-0.67$&$-0.69$ \cite{Buchmann:2009uqa}&\multicolumn{2}{|c|}{$-31.93$}\\
\cline{3-8}
&$G_M^{*2}\,R_\mathrm{SM}$&$-0.16$&$-0.23$ \cite{Buchmann:2009uqa}&$0.05$&$0.07$ \cite{Buchmann:2009uqa}&$-0.12$&$-0.16$ \cite{Buchmann:2009uqa}\\
\hline
\hline
\multirow{ 3}{*}{\rotatebox[origin=c]{90}{$\widetilde S_i$}}&all multipoles&\multicolumn{2}{|c|}{$35.24$}&\multicolumn{2}{|c|}{$0.00$}&\multicolumn{2}{|c|}{$35.24$}\\
\cline{3-8}
&$G_M^{*2}$&\multicolumn{2}{|c|}{$31.82$}&\multicolumn{2}{|c|}{$0.00$}&\multicolumn{2}{|c|}{$31.82$}\\
\cline{3-8}
&$G_M^{*2}\,R_\mathrm{SM}$&\multicolumn{2}{|c|}{$0.08$}&\multicolumn{2}{|c|}{$-0.03$}&\multicolumn{2}{|c|}{$0.05$}\\
\hline
\hline
\multirow{ 3}{*}{\rotatebox[origin=c]{90}{$S_i$}}&all multipoles&\multicolumn{2}{|c|}{$0.42$}&\multicolumn{2}{|c|}{$-0.71$}&\multicolumn{2}{|c|}{$-0.29$}\\
\cline{3-8}
&$G_M^{*2}$&\multicolumn{2}{|c|}{$0.55$}&\multicolumn{2}{|c|}{$-0.67$}&$-0.12$&$-0.12$ \cite{Faustov:1998mc}\\
\cline{3-8}
&$G_M^{*2}\,R_\mathrm{SM}$&\multicolumn{2}{|c|}{$-0.08$}&\multicolumn{2}{|c|}{$0.02$}&\multicolumn{2}{|c|}{$-0.06$}\\
\hline
\end{tabular}
\end{small}
\end{table}
\renewcommand{\arraystretch}{1.3}

\section{Summary and Conclusions}
The $\al^5$-proton-polarizability effect of the $\Delta(1232)$-resonance on the hydrogen spectrum is calculated from forward two-photon exchange, which in turn is related to the process of forward doubly-virtual Compton scattering off the proton. 

The main aim of this conference proceedings was to address a subtile difference between predictions based on a direct calculation of the two-photon exchange (or Compton scattering amplitudes) \cite{Faustov:1998mc} and predictions based on the $\Delta(1232)$-production photoabsorption cross sections \cite{Buchmann:2009uqa}. As we show in Section \ref{dif}, the tree-level Compton scattering with intermediate $\Delta(1232)$ exchange features terms with a structure of $\Delta$-pole type and a remainder, which we referred to as the ($\Delta$-)non-pole part. The former amplitudes can be reconstructed from the $\Delta(1232)$-production  photoabsorption cross sections with the help of dispersion relations. These cross sections feature a characteristic delta-function peaking at the $\Delta(1232)$-production threshold: $\delta(\nu-\nu_\Delta)$. For the latter non-pole amplitudes, we had to construct response functions by hand, cf.\ Eqs. \eref{npstrucfunc}, \eref{S2npg} and \eref{S2npgp2}. They are as well able to reproduce the non-pole amplitudes from dispersion relations, but have a different $\nu$- or $x$-dependence, respectively: $\delta(x)$.

Similar to Ref.~\cite{Buchmann:2009uqa}, we relate the $\Delta(1232)$ exchange to well-measured nucleon elastic form factors by means of the finite-momentum extension of the large-$N_c$ relations for the Jones--Scadron nucleon-to-delta transition form factors. While both $\Delta$-pole and non-pole amplitudes give substantial contributions to the Lamb shift and hyperfine structure, they largely cancel each other in the final result. Nevertheless, a study of the Burkhardt-Cottingham sum rule has proven that both $\Delta$-pole and non-pole amplitudes need to be considered, otherwise, the sum rule would be violated. We therefore claim that Ref.~\cite{Buchmann:2009uqa} agrees with our result for the $\Delta$-pole contribution, but misses the non-pole contribution to give a complete description of the effect of the $\Delta(1232)$-resonance on the hyperfine splitting in hydrogen.

The mentioned strong cancelations make the error estimate difficult and forced us to assign $100\,\%$ errors on Eqs. \eref{DeExLS} and
\eref{HFSmuH}. Also, they lead to the increased importance of $G_\mathrm{C}^*$ and $G_E^*$, as compared to the naturally dominating magnetic  transition $G_M^*$, cf.\ lower block of Table \ref{Comparison}.

To summarize, we have seen that our results for the hyperfine splitting of hydrogen agree very well with the literature \cite{Faustov:1998mc,Buchmann:2009uqa}. Furthermore, we have pointed out that it is important to consider the complete effect of the $\Delta(1232)$, including non-pole terms of the tree-level Compton scattering amplitudes, and that the electric and Coulomb Jones--Scadron form factors can not be neglected either.

\section*{Acknowledgements}

This work was supported by the Deutsche Forschungsgemeinschaft DFG through the
Collaborative Research Center SFB 1044 [The Low-Energy Frontier of the Standard Model] and the Graduate School DFG/GRK 1581 [Symmetry Breaking in Fundamental Interactions], and the Swiss National Science Foundation. Furthermore, it is a pleasure to thank Vladimir Pascalutsa for collaboration on some of the topics discussed in here, and Alfons Buchmann for useful discussions during the NSTAR 2017 conference which lead to the particular focus of the present proceedings. 
\appendix

\section{Tree-Level Compton Scattering with $\Delta(1232)$ exchange}
In the following, we present our results for the tree-level CS with intermediate $\Delta(1232)$ exchange, see \Figref{DeltaExchange}. The VVCS amplitudes are presented in the present section, whereas the associated $\Delta(1232)$-production cross sections, see \Figref{CrossSection}, are presented in the next section. In both sections, we make use of the following shorthand notations:
\begin{subequations}
\eqlab{shorthands}
\bea
\varDelta&=&M_\Delta - M,\\
M_+&=&M_\Delta + M,\\
\vert \bq \vert&=&\sqrt{\nu^2+Q^2},\\
Q_\pm&=&\sqrt{(M_\Delta\pm M)^2+Q^2}, \\
\omega_\pm&=&(M_\Delta^2-M^2\pm Q^2)/2M_\Delta.
\eea
\end{subequations}
\subsection{Compton Scattering Amplitudes} \label{amps}
Omitting the prefactor $\left[(s-M_\Delta^2)(u-M_\Delta^2)\right]^{-1}$, the $\Delta$-pole contributions read:
\begin{subequations}
\eqlab{Dpole}
\begin{alignat}{3}
&T_1^{\Delta\mathrm{-pole}}(\nu,Q^2)&&\propto\frac{2 \pi \al \nu^2 Q_-^2}{M M_\Delta M_+^2 \omega_+}\Big[g_M^2 Q_+^4+4 g_E^2 M_\Delta^2 \omega_-^2+4 g_\mathrm{C}^2 Q^4\\
&&&\quad-2 g_M g_E M_\Delta Q_+^2 \omega_-+2 g_M g_\mathrm{C} Q^2 Q_+^2-8 g_E g_\mathrm{C} M_\Delta Q^2 \omega_-\Big]\nn,\\
&T_2^{\Delta\mathrm{-pole}}(\nu,Q^2)&&\propto\frac{8\pi \al M_\Delta Q^2\omega_+}{M M_+^2}\Big[g_M^2 Q_+^2+g_E^2 Q_-^2+\frac{g_\mathrm{C}^2 Q^2 Q_-^2}{M_\Delta^2}\\
&&&\quad-2 g_M g_E M_\Delta \omega_-+2 g_M g_\mathrm{C} Q^2\Big]\nn,\\
&S_1^{\Delta\mathrm{-pole}}(\nu,Q^2)&&\propto-\frac{4\pi \al M_\Delta^2 \omega_+^2}{M M_+^2}\left[g_M^2 Q_+^2+\frac{ g_E^2  \omega_- \left(\varDelta ^2-Q^2\right)}{\omega_+}+\frac{2 \varDelta  g_\mathrm{C}^2 Q^4}{M_\Delta^2 \omega_+}\right.\\
&&&\quad-\frac{2 g_M g_E \left(M_\Delta M Q^2+\varDelta ^2 M_+^2-Q^4\right)}{M_\Delta\omega_+}+2g_M g_\mathrm{C} Q^2\left\{4-\frac{M \omega_-}{M_\Delta \omega_+}\right\}\quad\nn\\
&&&\quad\left.-\frac{2g_E g_\mathrm{C} Q^2 ( \omega_- (2 M_\Delta-M)- \varDelta   M)}{M_\Delta \omega_+}\right],\nn\\
&\left[\nu S_2\right]^{\Delta\mathrm{-pole}}(\nu,Q^2)&&\propto\frac{2\pi \al  M_\Delta^2 \omega_+}{M_+^2 M}\left[2 g_M^2 M Q_+^2+4 g_E^2 M_\Delta \varDelta \omega_-\right.\\
&&&\quad-\frac{2 g_\mathrm{C}^2 Q^2 \left(\varDelta ^2-Q^2\right)}{M_\Delta}+4 g_M g_E M_\Delta \left(M_\Delta \omega_+-4 M \omega_-\right)\nn\\
&&&\quad+\frac{g_M g_\mathrm{C} \left(16 M_\Delta M Q^2+\varDelta ^2 M_+^2-Q^4\right)}{M_\Delta}\nn\\
&&&\quad\left.+\frac{g_E g_\mathrm{C} \left(M_\Delta^4-6 M_\Delta^2 Q^2-M^4+2 M_\Delta M^3-2 M_\Delta^3 M+6 M_\Delta M Q^2+Q^4\right)}{M_\Delta}\right].\nn
\end{alignat}
\end{subequations}
The ($\Delta-$)non-pole contributions are given by:
\begin{subequations}
\eqlab{noDpole}
\bea
\widetilde T_1(\nu,Q^2)&=&-\frac{4\pi \al  \nu^2}{M M_+^2}\left[g_M^2+g_E^2-g_M g_E\right],\eqlab{T1nonpole}\\
\widetilde T_2(\nu,Q^2)&=&-\frac{4\pi \al Q^2}{M M_+^2}\left[g_M^2+g_E^2-g_M g_E+\frac{g_\mathrm{C}^2 Q^2}{M_\Delta^2}\right],\\
\widetilde S_1(\nu,Q^2)&=&\frac{\pi \al }{M M_+^2}\left[g_M^2 Q_+^2+g_E^2 \left(\varDelta ^2-3 Q^2\right)+\frac{4 g_\mathrm{C}^2 Q^4}{M_\Delta^2}-8 g_M g_E M_\Delta \omega_-\right.\eqlab{S1nonpole}\\
&&\qquad\qquad\left.-\frac{2 g_M g_\mathrm{C} Q^2 (M-4 M_\Delta)}{M_\Delta}+\frac{2 g_E g_\mathrm{C} Q^2 (3 M-2 M_\Delta)}{M_\Delta}\right],\qquad\nn\\
\widetilde{\nu S_2}(\nu,Q^2)&=&\frac{2\pi \al }{M M_+^2}\bigg[g_E^2 \,M_\Delta \varDelta\, \omega_-+\frac{g_M^2\, M Q_+^2}{2}+\frac{g_\mathrm{C}^2\,Q^2(Q^2-\varDelta^2)}{2M_\Delta}\eqlab{NUnonpoleS2}\\
&&\qquad\qquad+g_E g_M \,M_\Delta (M_\Delta \omega_+-4 M \omega_-)-g_E g_\mathrm{C}\, \varDelta (2Q^2+M \omega_+)\nn\qquad\\
&&\qquad\qquad+g_M g_\mathrm{C} \,Q^2(4M-\omega_+)\bigg]+\frac{\widetilde S_2(\nu,Q^2)}{\nu}\left[\frac{M_\Delta^2 \,\omega_+^2}{M^2}+\nu^2\right],\nn
\eea
\end{subequations}
with the non-pole contribution to $S_2$:
\beq
\widetilde S_2(\nu,Q^2)=-\frac{2\pi \al M \nu}{M_\Delta M_+^2}\big[g_M +g_E \big]g_\mathrm{C}\,,\eqlab{nonpoleS2}
\eeq
The $T_1$ subtraction function equals:
\bea
T_1(0,Q^2)&=&\frac{4\pi \al Q^4}{M_\Delta M_+ \omega_+}\left[\frac{g_M^2}{Q^2}-\frac{g_E^2 \varDelta}{M^2 M_+}-\frac{g_\mathrm{C}^2 \varDelta \left(M^2-Q^2\right)}{M^2 M_\Delta^2 M_+}+\frac{g_M g_E}{M M_+}\right.\eqlab{T1su}\\
&&\hspace{2.3cm}+\frac{g_M g_\mathrm{C}}{M M_+}+\left.\frac{2 g_E g_\mathrm{C} \left(M\varDelta +Q^2\right)}{M^2 M_\Delta M_+}\right].\nn
\eea

\subsection{$\Delta(1232)$-production Photoabsorption Cross Sections}\label{PhotoCS}
Here, we give our results for the proton structure functions:
\begin{subequations}
\eqlab{structurefunc}
\begin{alignat}{3}
&f_1(\nu,Q^2)&&=\frac{1}{2M_+^2}\left[g_M^2 \vert \bq \vert ^2 (\nu +M_+)+\frac{g_E^2\, (\nu -\varDelta ) \left(M\nu -Q^2\right)^2}{M^2}+\frac{g_\mathrm{C}^2 \,Q^4 s (\nu -\varDelta )}{M^2 M_\Delta^2}\right.\\
&&&\qquad\qquad-\frac{g_M g_E\, \vert \bq \vert ^2 \left(M\nu -Q^2\right)}{M}+\frac{g_M g_\mathrm{C}\, \vert \bq \vert ^2 Q^2}{M}\nn\\
&&&\qquad\qquad+\left.\frac{2 g_E g_\mathrm{C} \,Q^2 \left(M\nu -Q^2\right) (-M_\Delta(M+\nu)+s)}{M^2 M_\Delta}\right]\delta\!\left(\nu-\nu_\Delta\right),\nn\\
&f_2(\nu,Q^2)&&=\frac{\nu Q^2}{2 M M_+^2}\left[g_M^2 (\nu +M_+)+g_E^2 (\nu -\varDelta )-\frac{g_\mathrm{C}^2 \,Q^2 (\varDelta -\nu )}{M_\Delta^2}\right.\\
&&&\qquad\qquad\quad\left.-\frac{g_M g_E \left(M\nu -Q^2\right)}{M}+\frac{g_M g_\mathrm{C} \,Q^2}{M}\right]\delta\!\left(\nu-\nu_\Delta\right),\qquad\nn\\
&g_1(\nu,Q^2)&&=-\frac{\nu}{4M_+^2}\left[g_M^2 \nu  (\nu +M_+)+\frac{g_E^2 \left(\nu  M-Q^2\right) \left(M(\nu-\varDelta)-Q^2\right)}{M^2}\right.\eqlab{g1Delta}\\
&&&\qquad \qquad\quad-\frac{g_\mathrm{C}^2 \,Q^4 (M_\Delta M-s)}{M^2 M_\Delta^2} -\frac{g_M g_E \left(M_\Delta Q^2+4 \nu \left(M\nu-Q^2\right)\right)}{M}\nn\\
&&&\qquad \qquad\quad -\frac{g_M g_\mathrm{C}\, Q^2 \left(-4 \nu  M_\Delta+M \nu -Q^2\right)}{M M_\Delta}\nn\\
&&&\qquad \qquad\quad \left.+\frac{g_E g_\mathrm{C}\, Q^2 \left(\nu  M^2+\varDelta  \left(Q^2-  s\right)\right)}{M^2 M_\Delta}\right]\delta\!\left(\nu-\nu_\Delta\right),\nn\\
&g_2(\nu,Q^2)&&=\frac{\nu^2}{4M_+^2}\left[g_M^2 (\nu +M_+)+\frac{ g_E^2 \,\varDelta\left(\nu  M-Q^2\right)}{M^2}+\frac{g_\mathrm{C}^2 \,Q^2 (\nu  M_\Delta M-\varDelta  s)}{M^2 M_\Delta^2}\right.\eqlab{g2Delta}\\
&&&\qquad \qquad\quad-\frac{g_M g_E \left(-\nu  M_\Delta+4\left(M \nu  - Q^2\right)\right)}{M}\nn\\
&&&\qquad \qquad\quad-\frac{g_E g_\mathrm{C} \left(M \vert \bq \vert ^2\left(M-2 M_\Delta \right)+\nu  M_+ Q^2-Q^4+Q^2 s+\varDelta  \nu  s\right)}{M^2 M_\Delta}\nn\\
&&&\qquad \qquad\quad \left.+\frac{g_M g_\mathrm{C} \left(4 M_\Delta Q^2+\nu \left(M \nu-  Q^2\right)\right)}{M M_\Delta}\right]\delta\!\left(\nu-\nu_\Delta\right),\eqlab{g2Delta}\nn
\end{alignat}
\end{subequations}
cf.\ \Eqref{VVCSunitarity} for their relation to the $\Delta(1232)$-production photoabsorption cross sections.
Our results are conform with the helicity amplitudes of the $\Delta(1232)$-production mechanism presented in Ref.~\cite[Section 2.1.1.]{Pascalutsa:2006up}.




\end{document}